# Self-Organized Networks, Darwinian Evolution of Dynein Rings, Stalks and Stalk Heads

J. C. Phillips

Dept. of Physics and Astronomy, Rutgers University, Piscataway, N. J., 08854

Abstract

Cytoskeletons are self-organized networks based on polymerized proteins: actin, tubulin, and driven by motor proteins, such as myosin, kinesin and dynein. Their positive Darwinian evolution enables them to approach optimized functionality (self-organized criticality). Dynein has three distinct titled subunits, but how these units connect to function as a molecular motor is mysterious. Dynein binds to tubulin through two coiled coil stalks and a stalk head. The energy used to alter the head binding and propel cargo along tubulin is supplied by ATP at a ring 1500 amino acids away. Here we show how many details of this extremely distant interaction are explained by water waves quantified by thermodynamic scaling. Water waves have shaped all proteins throughout positive Darwinian evolution, and many aspects of long-range water-protein interactions are universal (described by self-organized criticality). Dynein water waves resembling tsunami produce nearly optimal energy transport over 1500 amino acids along dynein's one-dimensional peptide backbone. More specifically, this paper identifies many similarities in the function and evolution of dynein compared to other cytoskeleton proteins such as actin, myosin, and tubulin.

\body

**Introduction**

Long-range (allosteric) interactions often occur in proteins, and dynein is one of the most dramatic examples. Dynein is a large microtubule-based motor complex that requires tight coupling of intra-molecular ATP hydrolysis with the generation of mechanical force and track-binding activity. However, the microtubule-binding domain (MTBD) is structurally separated by about 20 nm and 1500 amino acids from the ATP nucleotide-binding sites. Thus, long-range two-way communication is necessary for coordination between the catalytic cycle of ATP hydrolysis and dynein's track-binding affinities [1,2]. The long range (cooperative) metabolic interactions of hemoglobin occur between hemes separated by 100 amino acids, and were long considered mysterious, but have recently been explained in terms of coupling of linear strain fields [3].

Extensive coarse-grained calculations of dynein interactions based on available structural data have shown that short-range contact interactions between different subunits of dynein probably contribute to the long-range connections between active ATP sites in the pre- and post-power cycle [4]. While the overall method is attractive, the two human-slime mold crystal structure sequences used in [4] have only 46% identity and 4% gaps, creating many uncertainties. Here we focus on evolution of dynein sequences with much larger identities (~ 60 - 76 %). Our analysis is restricted to the critical properties in only one dimension, and lacks the Euclidean contacts seen in structural models, but it is more accurate in quantifying evolutionary differences than standard coarse-grained models based on $C\alpha$ coordinates alone [4,5].

The active parts of dynein motors consist of three parts, the AA1-AA6 rings, the antiparallel coiled coil CC1 and CC2 stalks extended from AA4 and connected to the stalk head, and the CC buttress extended from A5. Each ring contains about 300 amino acids (aa), while CC1, CC2 and the stalk head contain about 100 aa each [4]. Here we use thermodynamic scaling to analyze the rings and the stalk, and identify their evolutionary refinements. Thermodynamic scaling utilizes recently discovered critical conformational features of hydropathic interactions which are associated with long-range strain-field interactions, and thus is well suited to discussing subtle evolutionary changes in shape connected with such long-range interactions.

Thermodynamic scaling utilizes only one-dimensional amino acid sequences, so it appears to lack the large quantity of information contained in three-dimensional static structures. Traditional methods for analyzing sequence evolution (phylogenetics) utilize single-site comparisons (BLAST and its derivatives). Single-site methods often lead to the conclusion that evolution is merely neutral and not progressive or positive [6]. One of the goals of thermodynamic scaling is to show the improvements evolution makes for many proteins, as discussed in earlier articles on cytoskeleton proteins (actin and tubulin [7,8]). Technically such improvement is expected for self-organized networks approaching a critical point [9,10].

Thermodynamic transitions involving short-range interactions, such as ATP hydrolysis, can be described as thermodynamically first-order, as can large changes, such as unfolding or cleaving. Long-range interactions are second order and small, and are simplified near a thermodynamic critical point, where a recently discovered hydropathicity scale (MZ) is extremely accurate in describing long-range water-mediated interactions [11]. Altogether 127 hydropathicity scales were proposed by 2000, which utilized small data bases and were never compared to each other, leaving the impression that they were all qualitative. The most popular of the early scales (KD) has turned out to be the second most accurate overall [12,13], and is effective in describing strong interactions, such as globin metabolism. The long range (cooperative) metabolic interactions of hemoglobin are quantified by combining MZ and KD evolutionary profiles [3].

**Results**

Dynein sequences are very well conserved (human and chicken, 98% identity). Given a huge choice of sequences from the genomic era data base, we have chosen to study evolution of dynein heavy chain (stem excluded) from worm (Q14204) to human or mouse (P34036), rather than from slime mold (A0A1S0UA63) to human. BLAST reports stronger identities for worm (68%) and positives (81%) than for slime (46%) and (65%), and only a few gaps (1% for worm, 4% for slime).

Our first step is to compare the hydropathic "roughness" or variance ratios of the hydropathic profiles Ψ(aa,W) as functions of the sliding window width W [3,7-10], calculated separately for the AAA1-AAA3 rings and the predominantly coiled coil stalk. The results using the short-range KD scale are shown in Fig. 1, and those for the MZ scale in Fig. 2. As expected, the curves with the two scales are roughly similar, and the stalk is qualitatively different from the rings. However, these important qualitative differences are absent for W = 1, which is why phylogenetics was unable to prove positive evolution from sequences alone [6,13].

The criteria that guide us in choosing a value of W = W* to display a profile Ψ(aa,W*) are that ideally it should be the smallest value (hence strongest resolution) which shows a peak in evolutionary variance ratios. In Fig. 2 we see such a peak for the stalk at W * = 13; it has a special significance, discussed below. With the KD scale a broader peak occurs for W* = 21, which also appears to be significant. The ring variance ratios do not show such a peak, but they do level off near W* = 29.

Dynein has two major ATPase sites in the AAA1 and AAA3 rings. ATP binding to AAA1 triggers a cascade of conformational changes that propagate to all six AAA domains, while nucleotide transitions in AAA3 gate the transmission of conformational changes between AAA1 and the stalk [14,15]. In Fig. 3 the worm and human profiles Ψ(aa,29) show a large qualitative difference in the hydrophilic minima separating AAA2 from AAA3. It seems likely that these hydrophilic extrema define the edges of these rings. In worm the three rings have similar sizes, and this is generally assumed for animals (for instance, on Uniprot "by sequence similarity"), but there are no structural data for animal species. Dynein fish sequences are incomplete, but frog dynein (F6XKW0) is similar to worm near the AAA2-AAA3 interface.

The deep human minimum in Fig. 3 arises from the strongly hydrophilic sequence 2396 RRRKGKEDEGEE 2407. This 12 aa sequence is conserved for almost all animals and birds. A possible explanation for the appearance of this extremely hydrophilic sequence is that it enhances the gating of AAA1 by AAA3. Note that most of the early (pre-2000) hydropathicity tables also list R, K, D and E as the most hydrophilic amino acids [11]. This is the case for the KD scale shown in Fig. 4; it also exhibits a strong hydrophilic minimum for the human, but not the worm, interface. The secondary minimum (near 650 in Fig. 3) which gives a clear-cut AAA2-AAA3 interface for worm with the 2007 MZ scale (Fig. 3) is absent for the 1982 KD scale (Fig. 4), a feature that favors the modern MZ scale.

Both full-length human [16] and slime [17] cytoplasmic dynein-1 have been studied by cryo-electron microscopy, with emphasis on the shaft and its complex motor activity. While planar rings are observable, individual AAA units are not, so that differences between AAA2 and AAA3 in human and worm dynein have not been resolved [17]. Given steady improvements in cryo-EM techniques, it may be possible to test this evolutionary prediction in the future.

According to Figs. 1 and 2, the best description for the evolution of the stalk is obtained with the MZ scale and W = 13. The resulting profile is shown in Fig. 5. The CC1 and CC2 coiled coils are labelled from the hydrophobic heptad repeats discussed below. Between them is the stalk head that binds to tubulin. Hydropathic profiling shows that the three regions are qualitatively different across all three species. CC1 is overall hydrophilic, while CC2 has stabilizing and two deep hydrophilic hinges near 250 and 315, hydrophobic peaks at both ends, separated by a stabilizing hydrophobic peak. The structure of the stalk head is discussed further below.

What is the meaning of these hydropathic coiled coil oscillations? Here the coiled coils are functioning as mechanical springs, tilting the stalk head to drive cytoskeleton cargo. Compressed helical pitches exclude water (hydrophobic extrema), while expanded helical pitches increase solvent accessible area, and increase water density (hydrophilic extrema). Hydropathic waves can be excited with lower energy, and as linear water film surface waves can travel along the dynein surface great distances, much as linear seismic water waves travel across oceans [19,20,21]. These linear shallow surface waves explain the remarkable distance covered by ATP excitation in AAA1 across a thousand amino acids to reach the stalk head. The more hydrophobic features of CC2 suggest that it can reflect the ATP wave from AAA1 back to the stalk head, increasing the intensity there, much like a nonlinear tsunami wave approaching shore [21]. Note that the role of gravity in shallow water waves is played in protein hydropathic waves by the van der Waals dispersion attraction between water and amino acids [22]. Also note that all our profiles are based on linear average; nonlinear tsunami effects could be even larger, but are beyond present methods.

The CC1 and CC2 coiled coils are stabilized by hydrophobic heptad repeats based on similar amino acids with one or two [1,2] $CH_3$ side groups and no benzenoid rings. [23]. These amino acids are (KD hydropathicities); hydroneutral Gly is 157 [1]): Ile (254),[2], Leu (240),[2], Val (248)[2], Ala (214)[1], and Met (202)[1]). The heptad repeats span 109 amino acids for both CC1 and CC2. The heptad sequence analysis [23] establishes KD as the scale better suited to describing short-range CC1-CC2 contact binding, because small Ala appears as hydroneutral on the MZ scale, which instead emphasizes long-range curvatures.

The KD heptad repeats appear to be similar for CC1 and CC2 [23], although the large scale MZ profiles with W = 13 are qualitatively different (Fig. 5). Looking more closely, we see that the hydrophilic minimum in CC1 near 70 corresponds to the hydrophobic maximum in antiparallel CC2 near 280. Binding CC1 to CC2 tends to compensate the variations in water density and thus supports hydrophobic heptad-mediated binding by reducing variations in helical pitches.

The most dramatic effects of dynein evolution occur in the stalk head, whose profile is shown in Fig. 6, enlarged from the central region of Fig. 5. Binding to tubulin occurs through the three hydrophobic peaks A-C. In slime mold Peak A is the strongest, and should provide most of the binding. In animals all three peaks can contribute to increasing the binding strength. This is in good agreement with experiment, which finds that animal stalk heads bind to tubulin in vitro, while slime and other primitive stalk heads do not [17,23].

More generally, evolution has leveled sets of (more often) hydrophobic and (less often) hydrophilic extrema in many proteins [24]. Such leveling optimizes protein hydrodynamics, in accordance with Sethian's level set hydrodynamic theory [25-27] Looking more closely at Fig. 6, we see that cargos in tubulin could be moved by rocking the shaft head between the two stronger peaks A and B, with peak C serving as a pivot. This corresponds to the pre- and post-stroke conformations observed with optical tweezers [28,29]. Note that some of the motor features are merely mechanical. For instance, the CC hinges near the shaft head are probably mechanical and not associated with the water film [17]. The CC in AAA5 may act primarily as a buttress supporting the shaft, as CC2 has already hydrophobically reflected the ATP AA1 water wave signal back to the shaft head. A model showing small tilts of flexible stalks observed by polarized total internal reflection fluorescence microscopy is also consistent with this rocking model [30].

**Discussion**

The complexity of the architecture of dynein and the difficulties in carrying out atomically detailed simulations, both due to the long time scales and inaccuracies in the force fields (especially water), have led to the creation of coarse grained (CG) models. Such models with multiple parameters based on crystal structures from several species have been used to analyze AAA and linker interactions separately [31]. The methods used here on dynein have previously been tested on other cytoskeleton proteins [7-9], as well as many other proteins [3,10,24], and have consistently yielded new insights into positive Darwinian evolution of protein functions from sequences alone. Because the genomic sequence data base is so much larger than all other protein data bases, in retrospect these multiple successes appear inevitable, once started. The starting problem here is associated with the "first-cut" nature of BLAST, which itself effortlessly yields many positive results. This has led many researchers to limit themselves to point (W = 1 or perhaps 2) comparisons of protein sequences, structure and function [6], without exploring the possibilities of optimizing wavelengths of hydropathic profiles by selecting larger values of W from evolutionary trends.

Here we have found that the positive effects of evolution are especially dramatic in dynein, where the linear waves in monolayer water films propagate along the protein chain 1500 amino acids from their source in AAA1 to the dynein stalk head tubulin binding sites. Within a W = 1 perspective, such "action at a distance" looks impossible in the presence of thermal fluctuations, while it is natural enough in terms of shallow water waves with W ~ 10-30 bound to an aqueously and critically self-organized amino acid backbone. Note that close to a critical point large density fluctuations may involve only small energy differences incorporated over long wavelengths.

Can the present results be extended using molecular dynamics simulations? Perhaps: the details of ATP hydrolysis have already been discussed for actin [32,7]. At least the rocking balance in the presence of thermal fluctuations between the A,B,C hydrophobic peaks in the dynein shaft head might be accessible. Comparisons of slime and human shaft head dynamics may be possible, and would test our picture of self-organized criticality.

An historical note: the underlying mechanism of molecular motors has been discussed for decades, for instance by Huxley (1957) and Feynman (1963), usually in terms of harnessing thermal fluctuations, a task apparently involving Maxwell demons for "thermal rectification" [33-35]. We have shown here that Darwinian selection has shaped motor proteins so that even linear water waves can transmit chemical energy over very long distances. Our statistical mechanical approach was anticipated by Schrodinger in 1943 [36,37]. Self-organized criticality was used in a schematic model to derive modular (or "punctuated") evolution in 1995 [38], and fractals [39] emerged explicitly in 2007 [11]. The concept of self-organized criticality, with potential applications to living matter, has been widely discussed for decades [40].

# References

1. Roberts, A. J.; Kon, T.; Knight, P. J.; et al. Functions and mechanics of dynein motor proteins. Nature Rev. Mol. Cell Biol. **14**, 713-726 (2013).

2. Rao, L.; Hulsemann, M.; Gennerich, A. Combining structure-function and single-molecule studies on cytoplasmic dynein. Single Mol. Anal.: Methods, Protoc.: Methods in Molecular Biology **1665**, 53-89 (2018).

3. Phillips, J. C. Thermodynamic scaling of interfering hemoglobin strain field waves. J. Phys. Chem. B **122**, 9324-9330 (2018).

4. Kubo, S.; Li, W.; Takada, S. Allosteric conformational change cascade in cytoplasmic dynein. PLOS Comp. Biol. 13, e1005748 (2017).

5. Wang, Q.; Jana, B.; Diehl, M. R.; et al. Molecular mechanisms of the interhead coordination by interhead tension in cytoplasmic dyneins. Proc. Nat. Acad. Sci. (USA) **115**, 10052-10057 (2018).

6. Nozawa, M.; Suzuki, Y.i; Nei, M. The neutral theory of molecular evolution in the genomic era. Ann. Rev Gen. Human Gen. **11**, 265-289 (2010).

7. Moret, M. A.; Zebende, G. F.; Phillips, J. C. Beyond phylogenetics I: Darwinian evolution of actin. Rev. Mex. Ingen.Biomed. **40**, 1-11 (2019).

8. Moret, M. A.; Zebende, G. F.; Phillips, J. C. Beyond phylogenetics II: tandem Darwinian evolution of tubulin. Phys. A **540**, 122886 (2019)

9. Bak, P.; Sneppen, K. Punctuated equilibrium and criticality in a simple model of evolution. Phys Rev Lett. **71**,4083-4086 (1993)..

10. Phillips, J. C Scaling and self-organized criticality in proteins: Lysozyme *c*. Phys. Rev. E **80**, 051916 (2009).

11. Moret, M. A.; Zebende, G. F. Amino acid hydrophobicity and accessible surface area. Phys. Rev. E **75**, 011920 (2007).

12. Kyte, J.; Doolittle, R. F. A simple method for displaying the hydropathic character of a protein. J. Mol. Biol. **157**, 105-132 (1982).

13. Philippe, H.; Brinkmann, H.; Lavrov, D. V.; et al. Resolving difficult phylogenetic questions: why more sequences are not enough. PLoS Bio. **9** (3): e1000602 (2011).

14. Bhabha, G.; Cheng, H.-C.; Zhang, N.; et al. Allosteric communication in the dynein motor domain. Cell **159**, 857-868 (2014).

15. Nicholas, M. P.; Berger, F.; Rao, L.; et al. Cytoplasmic dynein regulates its attachment to microtubules via nucleotide state-switched mechanosensing at multiple AAA domains. Proc. Nat. Acad. Sci. (USA) **112**, 6371-6376 (2015).

16. Zhang, K.; Foster, H. E.; Rondelet, A.; et al. Cryo-EM reveals how human cytoplasmic dynein is auto-inhibited and activated. Cell **169**, 1303-+ (2017).

17. Imai, H.; Shima, T.;[1]; Sutoh, K.; et al. Direct observation shows superposition and large scale flexibility within cytoplasmic dynein motors moving along microtubules. Nat. Commun. **6**, 8179 (2015)

18. Mallik, R; Carter, BC; Lex, SA; et al. Cytoplasmic dynein functions as a gear in response to load. Nature **427**, 649-652 (2004).

19. Camassa, R.;Holm, D. D. An integrable shallow-water equation with peaked solitons. Phys. Rev. Lett. **71**, 1661-1664 (1993).

20. Wang, Y.-L.; Gao, Y.-T.; Jia, S.-L.; et al. Solitons and integrability for a (2+1)-dimensional generalized variable-coefficient shallow water wave equation. Mod. Phys. Lett. B **31**, 1750012 (2017).

21. Constantin, A.; Henry, D. Solitons and Tsunamis. Z. Naturforsch. **64a**, 65-68 (2008).

22. Piana, S.; Donchev, A. G.; Robustelli, P.; et al. Water dispersion interactions strongly influence simulated structural properties of disordered protein states. J. Phys. Chem. B **119**, 5113-5123 (2015).

23. Gibbons, I. R.; Garbarino, J. E.; Tan, C. E.; et al. The affinity of the dynein microtubule-binding domain is modulated by the conformation of its coiled-coil stalk. J. Bio. Chem. **280**, 23960-23965 (2005).

24. Allan, D. C., Phillips, J. C. Evolution of the ubiquitin-activating enzyme Uba1 (E1). Phys. A **483**, 456-461 (2017).

25. Saye, R. I.; Sethian, J. A. The Voronoi implicit interface method for computing multiphase physics. Proc. Nat. Acad. Sci. (USA) **49**, 19498-19503 (2011).

26. Chiappori, P.-A.; McCann, R. J.; Pass, B. Multi-to one-dimensional optimal transport. Comm. Pure Appl. Math. **70**, 2405-2444 (2017).

27. Niu, X.; Vardavas, R.; Caflisch, R. E.; et al. Level set simulation of directed self-assembly during epitaxial growth. Phys. Rev. B **74**, 193403 (2006).

28. Kinoshita, Y.; Kambara, T.; Nishikawa, K.; et al. Step sizes and rate constants of single-headed cytoplasmic dynein measured with optical tweezers. Sci. Rep. **8**, 16333 (2018).

29. Rao, L.; Berger, F.; Nicholas, M. P.; et al. Molecular mechanism of cytoplasmic dynein tension sensing. Nature Comm. **10**, 3332 (2019).

30. Lippert, L. G.; Dadosh, T.; Hadden, J. A.; et al. Angular measurements of the dynein ring reveal a stepping mechanism dependent on a flexible stalk. . Proc. Nat. Acad. Sci. (USA) 114, E4564-E4573 (2017).

31. Goldtzvik, Y.; Mugnai, M. L.; Thirumalai, D. Dynamics of allosteric transitions in dynein. Structure **26**, 1664-+ (2018)

32. McCullagh, M.; Saunders, M. G.; Voth, G.A. Unraveling the mystery of ATP hydrolysis in actin filaments. J. Am. Chem. Soc. 136, 13053-13058 (2014).

33. Vale, R.D.; Oosawa, F. Protein motors and Maxwell's demons: does mechanochemical transduction involve a thermal ratchet? Adv Biophys, **26,** 97-134 (1990).

34. Vologodskii, A. Energy transformation in biological molecular motors. Phys. Life Rev. **3**, 119–132 (2006).

35. Rozenbaum, V. M.; Shapochkina, I. V.; Teranishi, Y.; et al. High-temperature ratchets driven by deterministic and stochastic fluctuations. Phys. Rev. E **99**, 012103 (2019).

36. Symonds, N. What is Life - Schrodinger influence on biology. Quart. Review Biolo. **61**, 221 (1986).

37. Holliday, R. Physics and the origins of molecular biology. J. Genetics **85**, 93-97 (2006).

38. Bak, P.; Paczuski, M. Complexity, contingency, and criticality. Proc. Natl. Acad. Sci. U S A. **92**, 6689–6696 (1995).

39. Mandelbrot, B. B. The fractal geometry of nature (Freeman, 1982).

40. Munoz, M. A. Colloquium: Criticality and dynamical scaling in living systems. Rev. Mod. Phys. **90**, 031001 (2018).

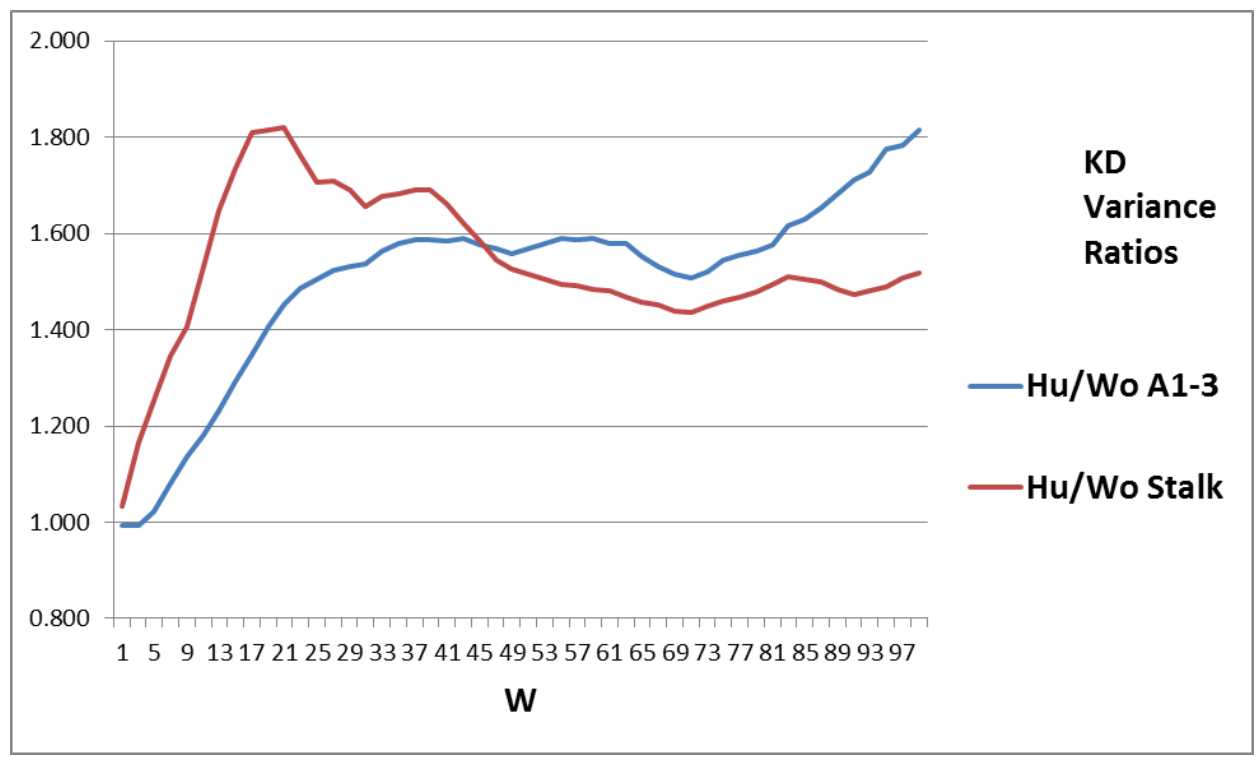

Fig. 1.  The human/worm variance ratios using the KD scale show a stronger peak near W = 21 for the stalk than for the AAA1-AAA3 rings.

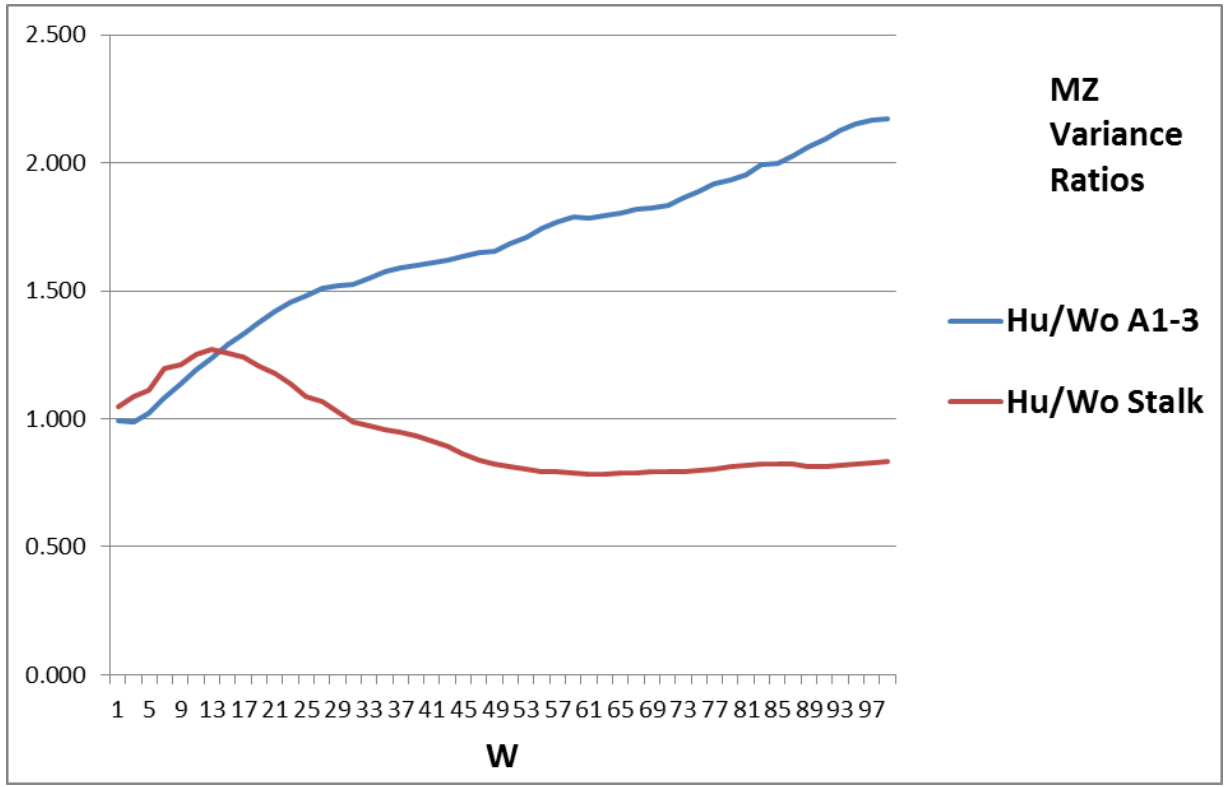

Fig. 2. We are again looking at the Human/Worm variance ratios With the MZ scale there is an unambiguous peak at W = 13 for the stalk, while the rings merely show a flattening near W = 29. The large differences from Fig. 1 (KD scale) reflect changes in short- vs. long- range forces.

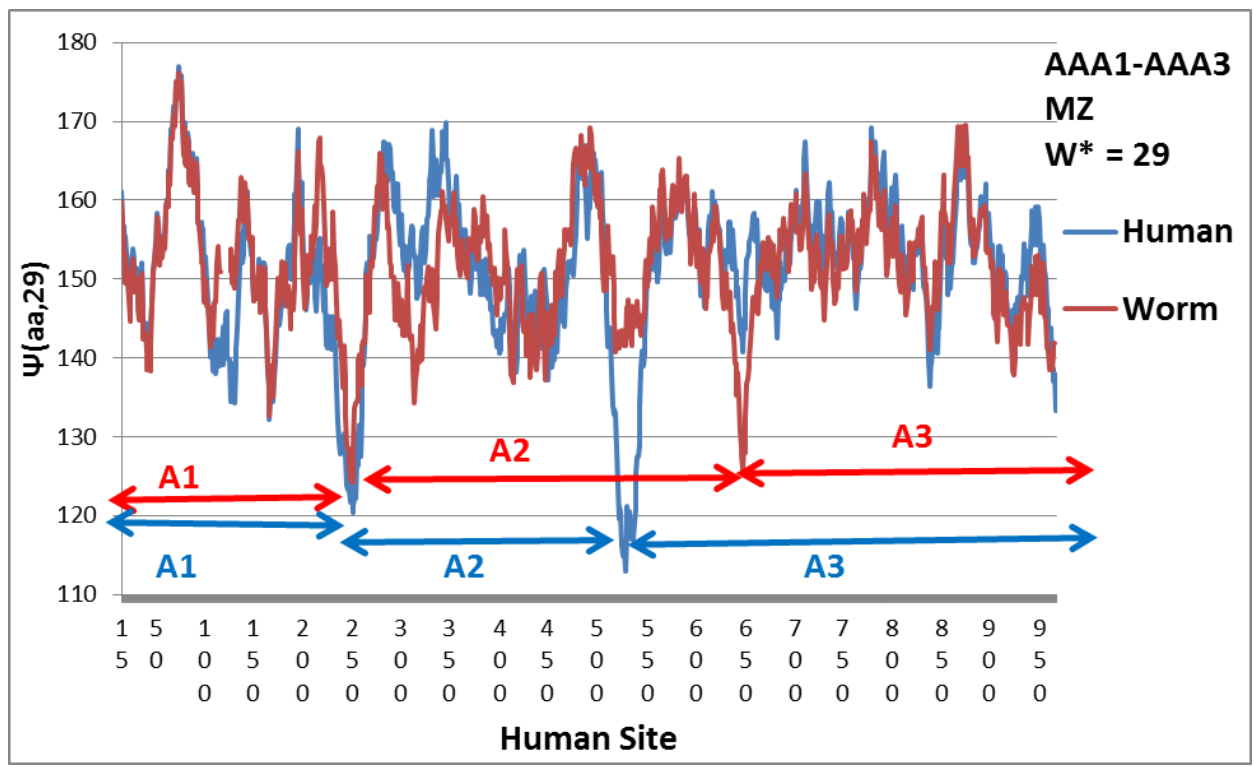

Fig. 3. Profiles of the human and worm rings. Here human site 1 corresponds to Uniprot Q14204 site 1868 Tyr. Worm is aligned to human by BLAST. A nine amino acid gap in worm is barely visible near 110. The largest change is the deep human hydrophilic minimum near 530, which improves decoupling of AAA2 from AAA3 (see text). It is absent from worm, which suggests that worm AAA2 and AAA3 may be decoupled near 635.

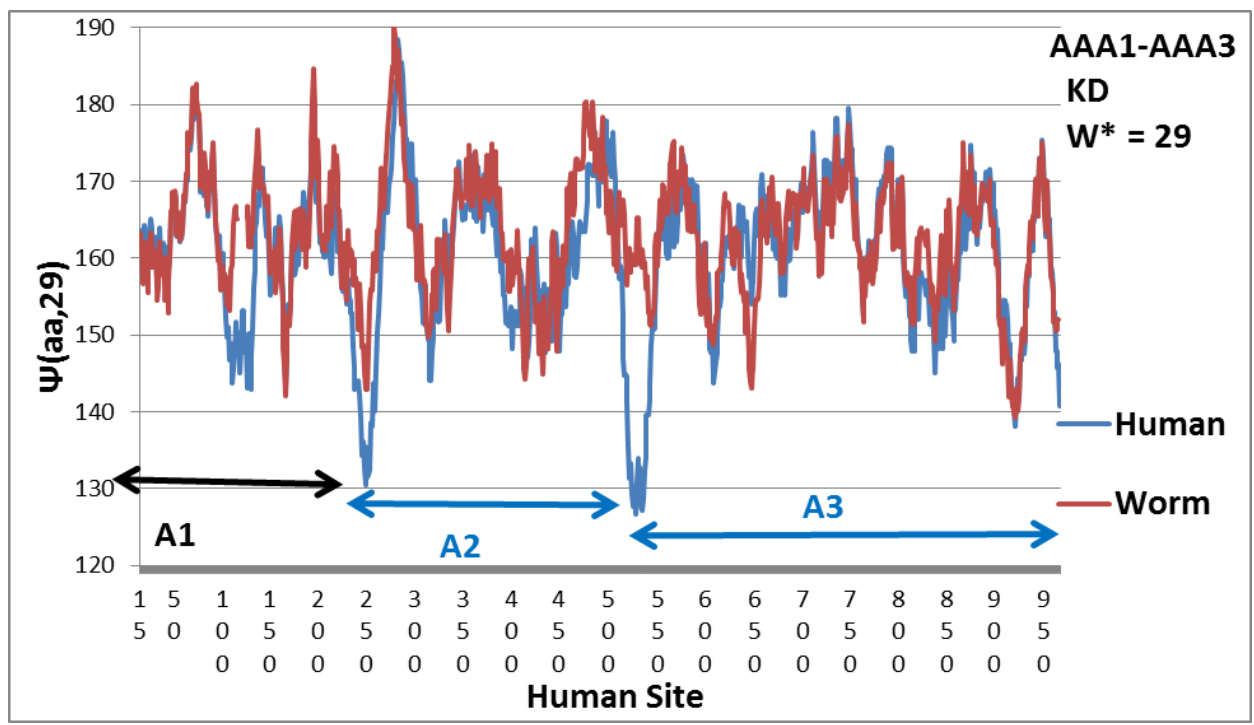

Fig. 4. With the KD scale, there is still a deep human hydrophilic minimum near 530 for human, but for worm there is no clear alternative decoupling of AAA2 from AAA3.

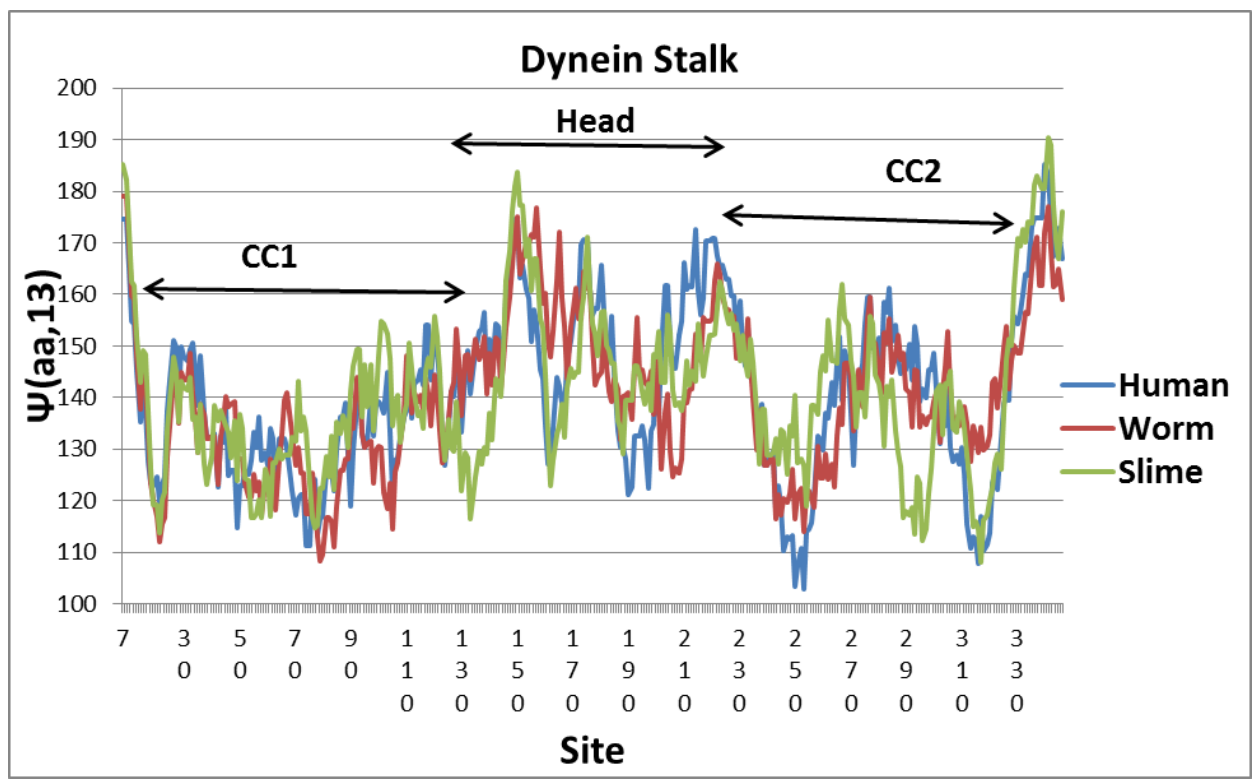

Fig. 5. Here site number 1 corresponds to Uniprot Q14204 site 3175 His). The regions labelled CC1 and CC2 are those where [19] identified heptad hydrophobic repeats. Guided by the Stalk peak in Fig. 2, W = 13 is used here.

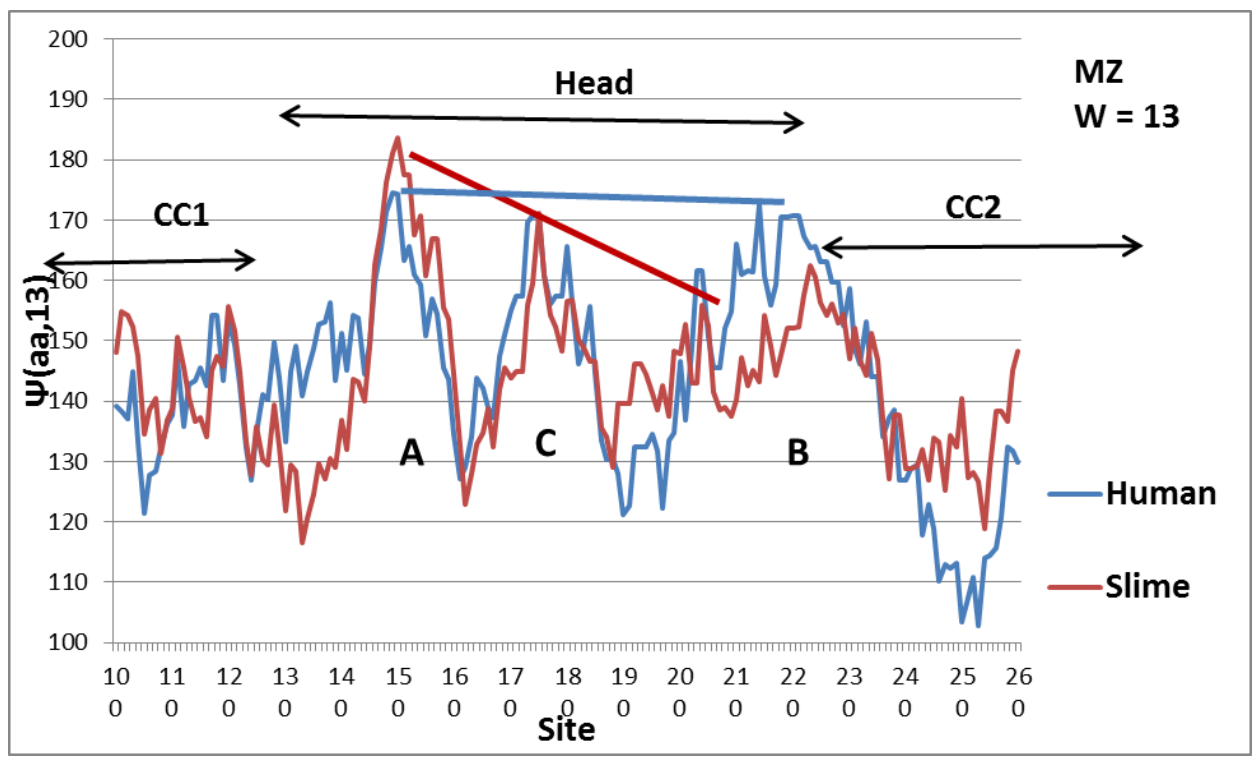

Fig. 6. An enlargement of the central region of Fig. 5. For clarity worm has been omitted. The three hydrophobic maxima A-C are tilted in slime mold, but are level in mouse and human. The CC hinges discussed in the text are located near 110 (CC1) and 240 (CC2). They appear to be mechanical, and are probably associated with the backbone amino acid packing.